\documentclass[10pt]{blois07}
\usepackage{graphicx}
\usepackage{cite,./mcite}

\usepackage{epsfig,multicol,amsmath}
\usepackage{wrapfig,rotating}

\def\beq{\begin{equation}}
\def\eeq{\end{equation}}
\def\bea{\begin{eqnarray}}
\def\eea{\end{eqnarray}}

\def\eq#1{{Eq.~(\ref{#1})}}
\def\fig#1{{Fig.~\ref{#1}}}
\newcommand{\bas}{\bar{\alpha}_S}
\newcommand{\as}{\alpha_S}

\newcommand{\Lb}{\left(}
\newcommand{\Rb}{\right)}

\parskip=\itemsep               

\begin{document}
\title{Non-linear QCD at high energies }
\author{E. ~Levin} 
\institute{
Department of Particle Physics, School of Physics and Astronomy,
Raymond and Beverly Sackler
 Faculty
of Exact Science  Tel Aviv University, Tel Aviv, 69978, Israel}

\maketitle
\begin{abstract}
In this talk I give the mini-review on  recent development in the non-linear QCD ( at low $x$).
\end{abstract}

This year is a jubilee   year: 25 years ago Leonid Gribov, Michael Ryskin and me pblished our GLR paper \cite{GLR} in which we set
a new field, so called,  high  parton density QCD or non-linear QCD.  In this paper we formulate the main physical question that we need to answer: what happens with the system of partons when their density becomes so large that the partons start to interact. This interaction was omitted in linear evolution but has to be important to  suppress the power like growth of the deep inelastic cross section which follows from linear evolution and contradicts the Froissart theorem.  The non-linear evolution equation which nowadays   goes under the name Balitsky-Kovchegov equation \cite{BK}  was suggested; the new scale: saturation momentum ($Q_s$), was introduced  and  the equation for this scale was derived; as well as the phenomenon of the saturation of the parton densities was  foreseen.

During the quater of century we have understood a lot: the role of the large number of colours in the approach \cite{BART}, a geometrical scaling behaviour of the scattering amplitude \cite{BALE,IIM}, the equation for the diffractive dissociation processes \cite{KOLE} and many other results. However, I think, we have had two major breakthroughs: the dipole approach \cite{MUCD}  and the colour glas condensate approach(CGC) \cite{MV,JIMWLK}. The dipole approach leads to a  new understanding 
what we calculate (dipole scattering amplitude), considerably  simplified all calculations and gives rise to statistical treatment of the problem.  CGC reduces the problem of  saturation to the theory of classical field in QCD giving the explanation of this phenomenon and developing a new theoretical method for the solution. I   firmly believe that we are now in the middle of the third breakthrough since we have started to attack the most difficult and challenging problem: the dynamical correlations in the QCD cascade which is known under slang name of summing Pomeron loops.  Therefore, the largest part of this talk I will spend on the discussion of this theoretical problem but I would like to start with more practical   question: are we ready for the LHC.

{\bf 1.  Practical impact on the LHC physics.}
The honest answer to the above question is firm no. I see two reasons for this sad fact: first the saturation physics is not  the hottest problem that the LHC hopes to resolve in spite of  having ALICE collaboration for  ion-ion collisions where the saturation effect will be more pronounced. Second, is a kind of contradiction between the theoretical approach and  the reality. Let me repeat 
what we are doing in hd QCD in more formal language. In the kinematic region where $\as\,\ln s \approx 1$ the asymptotical behaviour of QCD ampltude is known \cite{BFKL}  to be power-like as $A \propto  \as^2 s^{\Delta}$(BFKL Pomeron)   where $s$ is the energy and
$\Delta$ can be expanded as $\Delta = C_1 \as + C_2 \as^2$ with known coefficient $C_1$.  The calculation of $C_2$ has been performed \cite{NLOBFKL}  but these corrections  will be important only for $ \as \ln s \,\geq 1/\as$. 

 However, for lower energy , another type of interaction turns to be essential, namely, the exchange of two and more  BFKL Pomerons. Such exchange leads to  the contribution which is of the order of $\Lb \as^2  s^{\Delta_{LO}} \Rb^n$. Therefore when $ \as^2  s^{\Delta_{LO}}\,\,\approx\,\,1$ we need to sum all contributions due to BFKL Pomeron exchanges. In terms of energy this is the range $ 1 \leq  \as \ln s \leq 
\ln(1/\as^2)$. In simple words, theoreticaly first we need to solve the problem of high parton density QCD and only for higher energies we should take care about next-to-leading order corrections to $\Delta $. However, the life turns out to be more complicated and this corrections numerically are so huge that for any practical aplications we have to account them. The sad truth is we have not learned how to do this. As far as I know there is only one attempt to include them in non-linear evolution \cite{MODBK}  which is still very approximate.   It means that, frankly speaking, we cannot guarantee the value of the possible high parton density effects at the LHC. 

At the moment we can give  some estimates to illustrate how essential can be the high density collective effects at the LHC. The most important result  has been achieved during the past year. It turns out that the contribution of the semi-hard processes (  with the typical transverse momenta of the order of the saturation\ scale) to the   survival probability of the diffractive  Higgs  boson production is large
and it could lead to a substantial
 suppression of the QCD calculated cross section (see Refs. \cite{BBKM,JM}. 
 The estimates were obtained  for different contributions:  the fan diagrams in Ref. \cite{BBKM}  and 
 the enhanced diagrams in Ref. \cite{JM} with the same results. Namely, all diagrams of these types  should be summed. The model attempt to perform such a summation  with the result that the survival probability is as small as 0.4\%. This is a good example that we need to concentrate our efforts on LHC physics even in the case of the first wave of experiments, in particular in Higgs search. 
 
{\bf 2.  Statistical approach: its beauty and problems.}
Based on the probability interpretation of the non-linear equation in the dipole approach \cite{MUCD} we have tried to develop the more general statistical-like scheme that would  include the Pomeron loops (see 
review \cite{REVSTAT} and references therein). 
 The hope is to rewrite the QCD evolution equations including Pomeron loops in the form of Langevin equation ($\bigotimes$ stands for all needed integrations):
 \beq \label{LEQ}
 \frac{d N}{d \ln s}\,\,\,=\,\,\,\as K \bigotimes \Lb N - N^2 \Rb \,+\,\zeta\,\,\,\,\,\,\,\, \mbox{with} \,\,\,\,\,\,\,\langle|\zeta|\rangle = 0 \,\,\,
 \langle|\zeta\,\,\zeta|\rangle \,\,
\,\neq\,\,\,0
\eeq

In Ref. \cite{KLP} it was proven that in QCD we can obtain \eq{LEQ} but the form of $ \langle|\zeta\,\,\zeta|\rangle$  is so complicated that , I think, there is no chance of solving \eq{LEQ}. The attempts to solve \eq{LEQ} were made in the QCD motivated models with  a lot of assumptions. All these assumptions (especially
that impact parameter is much larger than the dipole sizes)  
  are such that we are loosing the possibility to calculate the Pomeron loops.  The main physical result from these models and statistical like approach is the violation of the geometrical scaling behaviour \cite{MMI}. I do not think we can trust this prediction.

\begin{wrapfigure}{r}{0.4\columnwidth}
\epsfig{file=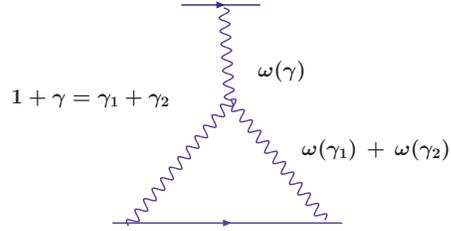,width=60mm}
\caption{ The first fan diagram. } \label{fan} 
\end{wrapfigure}

 {\bf 3.  BFKL Pomeron calculus: overlapping singularities.}
 The important news is the fact that
everything that has been done during the past three years is nothing more than understanding of the BFKL Pomeron calculus  \cite{KLP}.  Therefore we have to return back to this calculus  to re-examine  how to include the Pomeron loops in our  approach. In so doing in Refs. \cite{MUH,LMP} were found that the Pomeron interaction generates a new state with the intercept large than intercept of two BFKL Pomerons. In spite of the lack of room I will try to  illustrate  this result calculating the first fan diagram (see \fig{fan}).

 The expression for this diagram includes the integral over anomalous dimensions, dipole sizes in the triple Pomeron vertices which leads to  $\delta$ - function , shown in \fig{fan}, and the integral over $\omega$ which has the form:
 
   $ A\,\,\propto\,\,\frac{1}{2 \pi i}\,\int^{\epsilon + i \infty}_{\epsilon - i \infty}\,\,d \omega
e^{\omega\,Y}\,\frac{1}{\omega - \omega(\gamma)}\,\,\frac{1}{\omega\, -\, \omega(\gamma_1)\,-\, \omega(\gamma_2)}$

One can see that we cam close the contour over $\omega$ on two poles: $\omega(\gamma)$ and $\omega(\gamma_1)\,+\, \omega(\gamma_2)$, which lead to contribution $\exp \Lb \omega(1/2) Y\Rb$ and $\exp\Lb 2 \omega(1/2) Y\Rb$. However,  in the integral over $\gamma$ there exists $\gamma=\gamma_0$ which is a solution to the equation: $ \omega( 2\gamma_0 \,-\,1)\,\,=\,\, \,2\,\,\omega(\gamma_0)$. For $\gamma_0$ we have a double pole and the amplitude behaves as $ A\,\, \propto\,\,\,Y\,e^{2\,\,\omega(\gamma_0)\,Y}
\,\gg\,\,\exp \Lb \omega(1/2) Y\Rb$ . This new singularity we call overlapping singularity. Its appearance is kind of disaster since it means that even in 'fan' diagrams the partons from different parton showers, which described by exchange of two Pomerons ,  interacts. In particular,  overlapping singularities destroy the non-linear Balitsky-Kovchegov equation even for the scattering with nucleus, preserving nevertheless the Balitsky 
chain of equations\cite{BK,LELU}. So, the  truth is that we have to start from the beginning  not only in summing the Pomeron loops but also in the mean field approximation.

\begin{boldmath}
{\bf 4. BFKL Pomeron calculus: solution for $\as\, \,lns \,< \,1/\as$.}
\end{boldmath}
Therefore, the first thing that we need to do is to suggest our way to overcome the difficulties related to the overlapping singularities. 

\begin{figure}[h]
\begin{tabular}{c}
\epsfig{file=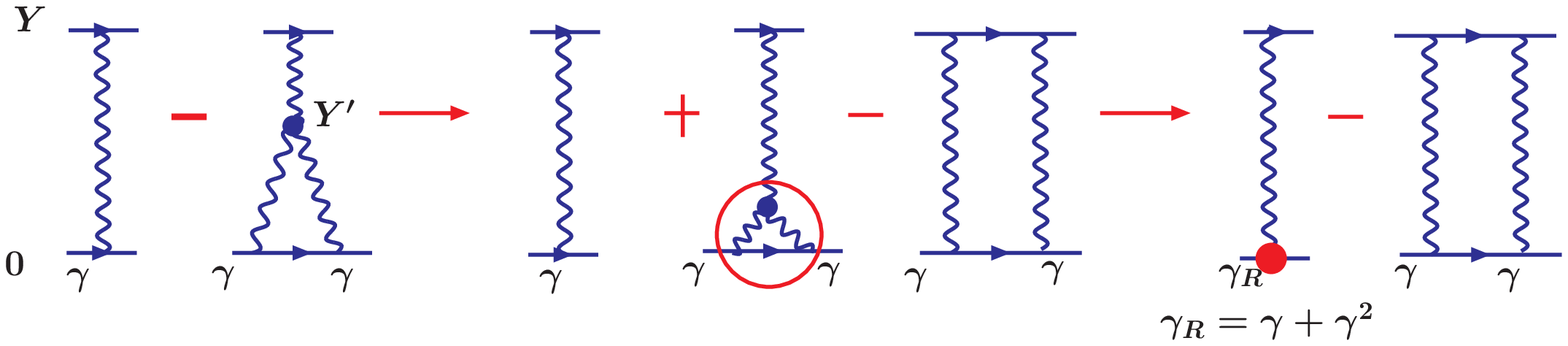,width=140mm,height=30mm}\\
\epsfig{file=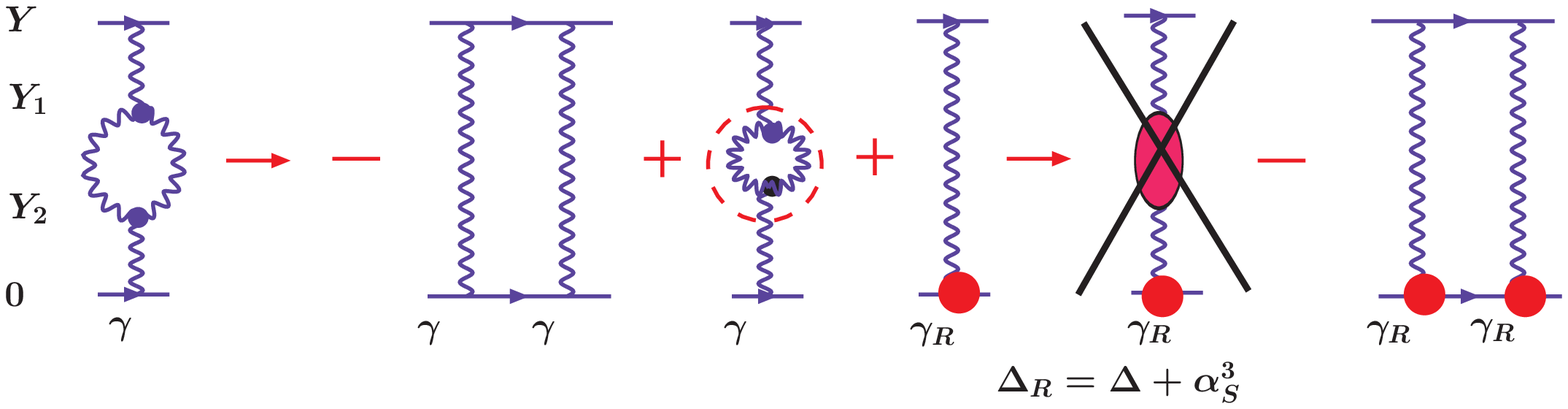,width=140mm,height=30mm}\\
\end{tabular}
\caption{The calculation of the first fan and enhanced diagrams.}
\label{dia}
\end{figure}

Our observation is the following:
  $\gamma_0 \,\,> \,\,\gamma_{cr}$ therefore, two Pomerons ($\gamma_1$ and $\gamma_2$) are inside the saturation region while one upper Pomeron is outside. Inside the saturation region we cannot use the BFKL
  kernel to determine $\omega(\gamma)$ but we need to use the expression found in Ref. \cite{BALE}, namely,
$\omega_{sat}(\gamma)\,\,=\,\,\frac{\omega(\gamma_{cr})}{1 - \gamma_{cr}}\,( 1 - \gamma)$. Noticing that
equation
$ 2 \omega_{sat}(\gamma_0)\,\,=\,\,\omega_{pert}(2\gamma_0 - 1)$ has no solution, we can conclude that 
as the first try we can neglect the overlapping singularities. However, we have to check the self consistence of our approach, namely, obtaining a solution to calculate back the diagrams and show that they indeed give a small contribution.  

Our main idea\cite{LMP}  that in this case and for
the kinematic region $\as\,ln s \,<\,1/\as$ we are dealing  with the system of non-interacting Pomerons,  \fig{dia}, in which we present the calculation of the first fan and the first enhanced diagrams,  illustrates this idea.   One can see in \fig{dia} that these diagrams can be reduced to the system of non-inetracting Pomerons since in the kinematic region under consideration the corrections to the Pomeron intercept turns out to be small. (see \fig{dia}). Having this fact in mind  we can use for summing  Pomeron loops the Iancu-Mueller-Patel -Salam approximation\cite{MPSI}, improved by the renormalization  of the scattering  amplitude at low energies. This approach is nothing more than the t-channel unitarity constraint adjusted to the dipole approach. 

However, the first part of the problem: to find the sum of fan diagrams we need to solve using a different method. We were able to do this for the simplified BFKL kernel in which we took into account only the leading twist part of the full BFKL kernel.  We heavily use the fact that in Ref.\cite{LT} the solution for this kernel have been found. The simplified kernel looks as follows 
\beq \label{SKE}
~\omega(\gamma )\,\,=\,\,\bas\,\,\left\{\begin{array}{c}
\,\,\,\,\,\,\frac{1}{\gamma}\,\,\hspace*{1cm}\mbox{for}\,\,r^2\,Q^2_s\, \ll\,1\,  \,\, \mbox{summing} \,\,\,\,(\bas \ln(1/(r^2\,Q^2_s)))^n; \\
~\\
\,\frac{1}{1\,-\,\gamma}\hspace*{1cm} \mbox{for}\,\,r^2\,Q^2_s\, \gg\,1\, \,\,\mbox{summing} \,\,\,(\bas
\ln(r^2\,Q^2_s))^n;
\end{array} \right.\\
\eeq
For this kernel we obtain: the solution that icludes the Pomeron loops,  with the following main properties:
,geometrical scaling behaviour and rather slow  approaching the asymptotic value, namely $1 - N\,\,\propto\,\,\exp( - z )$
where $z\,\,=\,\,\ln(r^2Q^2_s)$.

{\bf Resume} One my friend, a good experimentalist, told me, that what I am doing,   is the same as string theorists are doing : the approach is complicated and a lot of promises but no delivery( no connection with the reality) . I agree with him that the problem is not simple and during the last 25 years we learned how difficult  it is. However, the main difference with the string theory is that we are solving a well formulated theoretical problem about the nature while string theory is dealing with the imaginary world without any chance to approach reality and in attempts to include the  principle property of the nature they have to build models for each well establish phenomena:
running $\as$; confinement of quarks and gluons; and the violation of chiral symmetry. Accepting the fact that we have not prepared yet the experimental program for the LHC for measuring saturation effect we  
are developing  fast in this direction and I firmly believe that I will report very soon  to my friend 
about such program beating his second claim.

\begin{footnotesize}

\end{footnotesize}

\end{document}